\documentclass[prd,superscriptaddress,nofootinbib,11pt, tightenlines]{revtex4-2}
\usepackage{epsfig}
\usepackage{amsmath}
\usepackage{amsfonts}
\usepackage{xfrac}
\usepackage{amssymb}
\usepackage{slashed}
\usepackage{color}
\usepackage{pbox}
\usepackage{booktabs}
\usepackage{array,multirow,makecell}
\usepackage[colorlinks,citecolor=blue, linkcolor = blue, urlcolor = blue]{hyperref}
\usepackage{tabularx}
\usepackage{titlesec}
\usepackage{natbib}
\usepackage{lipsum}
\usepackage[normalem]{ulem}
\usepackage{xcolor}
\usepackage{marginnote}
\usepackage{cancel}
\usepackage{tikz}
\usepackage{tikz-feynman}
\usetikzlibrary{arrows.meta}
\tikzfeynmanset{compat=1.1.0}
\usepackage{enumitem}


\def\bea{\begin{eqnarray}}
		\def\eea{\end{eqnarray}}

\newcommand{\mdp}{M_{Z'}}

\newcommand{\AddrIOP}{
	Institute of Physics, Sachivalaya Marg, Bhubaneswar, 751 005, India}
\newcommand{\AddrIMSc}{
	Institute of Mathematical Sciences, Taramani, Chennai 600 113, India}
\newcommand{\AddrHBNI}{Homi Bhabha National Institute, Training School Complex, Anushakti Nagar, Mumbai 400 094,
	India}
\begin{document}
\title{SN1987A Constraints of Light $\boldsymbol{Z'}$ with Non-Mixing Polarisations
}

\author{Debottam Das}\email{debottam@iopb.res.in}
\affiliation{\AddrIOP}
\affiliation{\AddrHBNI}
\author{Purusottam Ghosh}\email{pghoshiitg@gmail.com}
\affiliation{\AddrIMSc}
\affiliation{\AddrHBNI}
\author{Rahul Puri}
\email{rahulgpuri@gmail.com}
\affiliation{\AddrIOP}
\affiliation{\AddrHBNI}

\begin{abstract}
	The observation of supernova 1987A (SN1987A) provides a unique opportunity to explore new physics beyond the Standard Model (BSM). The production of new particles in the supernova core could accelerate the cooling process, leading to additional energy loss and consequently reducing the duration of the observed neutrino burst at detectors. Therefore, any BSM interactions that affect supernova cooling are subject to stringent constraints from SN1987A observations. In this paper, we revisit the constraints on light gauge bosons (LGBs) by reassessing the validity of underlying assumptions about the polarisation intermixing. 
    We argue that the intermixing between different polarisation modes is suppressed in the low coupling regime.
    Using the light gauge boson in the $L_\mu-L_\tau$ model as an example, we find that considering the independent energy transport of longitudinal and transverse polarisations can lead to significant modifications of the SN1987A bounds on the parameter space.
\end{abstract}

\maketitle
\section{Introduction}
The core-collapse supernova 1987A (SN1987A)~\cite{Bionta:1987qt,Kamiokande-II:1987idp}, resulting from the explosion of the massive star Sanduleak $-69^\circ$ 202 in the Large Magellanic Cloud, may serve as an astrophysical laboratory for exploring new physics (NP) beyond the Standard Model (BSM) \cite{Kadota:2014mea,Chang:2022aas,Fiorillo:2024upk}. 
A core-collapse supernova (CCSN) is a huge explosion that occurs when a massive star runs out of nuclear fuel and loses its ability to resist the gravitational pull. 
This triggers the collapse of the iron core under gravity, producing super-nuclear density ($\rho \sim 3 \times 10^{14}$ g/cm$^3$) and hot temperature ($T \sim 30$ MeV) in the core. 
During core-collapse, electrons are captured by protons, increasing the number density of neutrons in the core and producing neutrinos: $p + e \to n + \nu_e$, known as neutronisation \cite{Janka:2006fh}. 
These neutrinos initially interact with the surrounding matter and are trapped in the dense core. 
However, as the density decreases outward from the core, neutrinos begin to escape. They continue to escape from the hot, newly formed proto-neutron star (PNS), carrying away energy and thus helping it cool over time. The supernova formed a PNS with mass $M_{\rm PNS} \sim 1.4 M_\odot$ and initial radius $R_{\rm PNS} \sim 25$ km. 
In 1987, neutrino detectors such as Kamiokande, Irvine-Michigan-Brookhaven (IMB) first detected a burst of neutrinos lasting about $10$-$12$ seconds coming from SN1987A~\cite{Bionta:1987qt,Kamiokande-II:1987idp}. 
These neutrinos were detected about 2-3 hours before visible light from the explosion reached Earth. 
These observations confirmed the theoretical expectation that the vast majority of the gravitational binding energy released during core collapse, $E_G \sim 3\times10^{53}\ \mathrm{erg}$, is emitted in the form of neutrinos~\cite{Sung:2019xie}. Indeed, nearly $99\%$ of the total energy liberated in the explosion is carried away by neutrinos. The duration and energetics of the observed neutrino burst are broadly consistent with Standard Model predictions, in which the cooling of the PNS proceeds through the emission of all six neutrino and antineutrino species~\cite{Janka:2006fh}.
In a supernova environment, electrons and electron-neutrinos dominate the leptonic content~\cite{Janka:2006fh}. Nevertheless, muons can be efficiently produced in the dense core due to the high temperature and large electron chemical potential ($\mu_e\gtrsim m_\mu$), via both pair production and the processes involving neutrinos. In contrast, tau production is strongly suppressed, since the core does not attain chemical potentials or temperatures sufficient to overcome the large tau mass threshold. For details on muon creation inside a supernova, see Ref.~\cite{Bollig:2017lki}.

The observation of the nearby supernova SN1987A provides a powerful astrophysical laboratory for testing and advancing our understanding of new physics, including those involving axion-like particles (ALPs), light gauge bosons or dark photons, sterile neutrinos, and other light exotic particles. 
The new long-lived particles produced inside the PNS can accelerate its cooling by carrying away energies comparable to the gravitational binding energy $E_G$, thus shortening the observed neutrino burst.
This refers to a constraint on energy loss due to the production of BSM particles inside a PNS, which should not exceed $3\times 10^{52}$ erg s$^{-1}$,  a limit known as the Raffelt criterion \cite{Raffelt:1996wa}. In the literature, this is typically referred to as SN cooling.
Based on this cooling argument, several BSM physics scenarios have been studied in the literature, including the models involving dark photons \cite{Chang:2016ntp,Sung:2019xie}, light gauge bosons originated from $U(1)_{X}$ extensions with $X=B-L$, $L_{\mu}-L_{\tau}$ \cite{Croon:2020lrf,Lai:2024mse, Akita:2023iwq, Blinov:2025aha}, ALPs \cite{Lucente:2020whw, Hook:2021ous,Lee:2018lcj, Croon:2020lrf}, sterile neutrinos \cite{Chauhan:2023sci,Carenza:2023old}, and light dark matters (DMs) \cite{Manzari:2023gkt,Sung:2021swd,Cappiello:2025tws}.
In addition to affecting SN cooling, the BSM particles produced inside the PNS might decay into photons or $e^\pm$ outside the progenitor star ($R_* \sim 10^9$ km), producing observable $\gamma$-rays/ X-rays \cite{Sung:2019xie,Jaeckel:2017tud,DeRocco:2019njg,Carenza:2023old}.
Non-observation of such signatures associated with BSM physics from SN1987A has also been used to constrain the BSM scenarios, complementing the cooling bounds. If these BSM particles decay inside the progenitor ($r < R_{*}$), it can lead to substantial energy deposition inside the supernova envelope ($R_{\rm PNS} < r < R_{*}$), potentially altering the explosion dynamics \cite{Caputo:2022mah}.
This argument, based on energy deposition, imposes an upper limit of $\sim 10^{50}$ ergs on the total energy or integrated luminosity, to avoid producing overly energetic explosions that would be inconsistent with observations \cite{Caputo:2022mah,Carenza:2023old,Fiorillo:2025sln}. 

In earlier studies involving the production of a dark photon (DP)~\cite{Croon:2020lrf, Lai:2024mse,Hardy:2016kme,Caputo:2021rux,Rrapaj:2015wgs}, the polarisation modes are assumed to intermix readily, preventing decomposition of the Boltzmann equation into separate equations for each polarisation. In such a scenario, a longitudinal (transverse) light gauge boson (LGB) can convert to a transverse (longitudinal) mode and be reabsorbed through transverse (longitudinal) channels.
While the mixing between the polarisation modes can be facilitated by scattering processes such as the \emph{dark-Compton} process ($\mu ~Z^\prime_L \leftrightarrow \mu ~Z^\prime_T$), they are suppressed by an additional factor of ${g^\prime}^2$ (where $g'$ is the new gauge coupling), compared to production and absorption rates of the semi-Compton process ($\mu ~\gamma \leftrightarrow \mu ~Z^\prime$). Such mixing modes (between the longitudinal and transverse polarisation) can therefore be neglected for appropriately small couplings. As a consequence, the coupled Boltzmann equation can be decomposed into an independent set of equations, one for each polarisation. Given the difference in the interaction rates for these polarisations, we show that such a decomposition modifies the SNe bounds, especially in low LGB mass regions.

In this context, we revisit the SN cooling constraints on a light gauge boson $Z^\prime$ originating from a $U(1)_X$ model that interacts with muons, taking into account the polarisation effects of $Z^\prime$. 
In particular, we focus on the $U(1)_X$ scenario with $X=L_\mu-L_\tau$, where the gauge boson couples only to the second and third generation of leptons. 
However, it can indirectly link with other SM fermions through loop-induced processes or kinetic mixing with $\gamma$. 
Within the supernova environment, $Z^\prime$ could be dominantly produced and reabsorbed through the pair coalescence process $\nu \bar\nu , ~\mu^+\mu^- \leftrightarrow Z^\prime$, semi-Compton processes $\mu~\gamma \leftrightarrow \mu ~Z^\prime$, and loop-Bremsstrahlung processes $n~p \leftrightarrow n~p~Z^\prime$. We revisit the production of $Z^\prime$ inside the SN core, taking into account polarisation effects. The longitudinal and transverse polarisation modes transport energy independently, together contributing to the total energy transport. We find some differences with existing studies in the SN cooling constraints on light gauge bosons $Z^\prime$~\cite{Croon:2020lrf, Lai:2024mse}. These differences may arise from the independent treatment of the polarisation. To the best of our understanding, this work presents a complete and consistent treatment of the underlying processes.

The rest of the paper is organised as follows. In Sec-\ref{sec:Improv}, we highlight the key features that distinguish our study from earlier works. 
Sec-\ref{sec:Prod} summarises the LGB production mechanism. In Sec-\ref{sec:DPModel}, we describe the BSM model that we use as an example, and calculate expressions for the production and absorption rates. Our results are presented in Sec-\ref{sec:Res}. Finally, we summarise our findings and conclude in Sec-\ref{sec:SnC}.


\section{Improved Treatment of Light Gauge Boson Emission in Supernovae}\label{sec:Improv}
New physics can facilitate the production of LGB in the PNS core due to the high temperature and density. 
If these LGBs are long-lived particles (LLP), \emph{i.e.} for sufficiently weak couplings to SM particles, they might be produced and escape the core, carrying energy outside the neutrinosphere ($R_\nu \sim 20$ km), providing a new channel for the PNS core cooling. 
Such an energy loss can modify the cooling dynamics of the supernova by reducing the observed and expected cooling duration. 
However, observations of neutrinos from SN1987A are consistent with simulations that consider neutrino emission as the sole cooling mechanism~\cite{Burrows:1986me,Burrows:1987zz}.
This agreement constrains any additional energy-loss mechanisms. In particular, the luminosity carried away by any new particle must not exceed the neutrino luminosity, as encapsulated by Raffelt’s criterion~\cite{Raffelt:1996wa}:
\begin{align}
L_{Z'} < L_\nu \simeq 3 \times 10^{52}~\mathrm{erg/s}~.
\end{align}
Consequently, the contribution of LLPs must remain subdominant, allowing one to place constraints on their coupling and mass.

Several works have attempted to estimate the luminosity of a new vector boson particle within the supernova context.  In this work, we revisit earlier estimates of the luminosity of LGBs and provide an improved analysis using a refined numerical approach, as outlined below:

\begin{itemize}
    \item 
    Here, we consider that the LGB polarisations do not intermix in the plasma for low couplings to SM particles. This is in contrast to earlier works~\cite{Croon:2020lrf, Lai:2024mse,Hardy:2016kme,Caputo:2021rux,Rrapaj:2015wgs}\footnote{While some of the references have evaluated separate rates for different polarisations, it is unclear whether the luminosity for each polarisation is evaluated independently.}, where the treatment of polarisations requires efficient polarisation intermixing.

   An LGB produced in the PNS core can subsequently interact with the ambient plasma via scattering, get reabsorbed, or escape. The collision terms in the Boltzmann equation can thus be decomposed as,
    \begin{align}
        (\partial_t+\partial_s)f^\lambda=C[f^\lambda]\equiv C_{\rm prod}[f^\lambda]+C_{\rm abs}[f^\lambda]+C_{\rm scat}[f^\lambda]+\dots~~~~,
    \end{align}
    where $\partial_s$ is the partial derivative with respect to the line element along the direction of propagation and $\partial_t$ is the time derivative. The scattering process such as the \emph{dark-Compton} ($\mu~ Z'\leftrightarrow\mu ~Z'$) is suppressed by an additional factor of $g'^2$ relative to the production and absorption processes (\emph{e.g.} semi-Compton) for appropriately small gauge coupling $g'$.
    Consequently, the corresponding scattering rate is subdominant, and the longitudinal-transverse collision term can be neglected in the Boltzmann equation, as follows from the following equation. 
\begin{align}
    (\partial_t+\partial_s)
        f_{\lambda}
    &=
    \underbrace{
            \vphantom{\sum_\kappa}\Gamma_{\rm prod}^\lambda f_\lambda^+-\Gamma^\lambda_{\rm abs}f_\lambda
    }_{\sim~\mathcal{O}(\alpha')}+
    \underbrace{
            \sum_\kappa\left[\Gamma_{\kappa\rightarrow\lambda } f_\lambda^+-~\Gamma_{\lambda\rightarrow \kappa}f_\lambda \right]
    }_{\sim~\mathcal{O}(\alpha'^2)}+\dots~~,\nonumber\\[5pt]
    (\partial_t+\partial_s)
        f_\lambda
    &\approx
        \Gamma_{\rm prod}^\lambda f_\lambda^+-\Gamma^\lambda_{\rm abs}f_\lambda~~,
\end{align}
where $f_\lambda^+=1+f_\lambda$ is the Bose enhancement factor for the polarisation $\lambda$ ($\lambda, \kappa \in \{L, T\}$, $T$ and $L$ being the transverse and longitudinal modes, respectively) and $\alpha^\prime={g^\prime}^2/4\pi$. As a result, the luminosity associated with each polarisation can be computed independently in the small coupling limit by solving separate Boltzmann equations. The total luminosity due to the emission of LGBs is then obtained by summing over the contributions from the individual polarisation states as,
\begin{align}
L_{Z'} = 2\,L_{Z'}^{T} + L_{Z'}^{L}\, .
\end{align}
The factor of 2 accounts for the two transverse degrees of freedom.
     \item The independent treatment of polarisation modes significantly affects the total luminosity. For lower LGB masses, where the semi-Compton processes dominate the $T$-mode production, the contribution of these modes ($2 L_{Z'}^{T}$) is more prominent at smaller couplings. 
     However, at larger couplings, the contribution from transverse modes decreases, while longitudinal modes ($L_{Z'}^{L}$) begin to contribute significantly and eventually dominate. As a result, for a given mass, two maxima appear at lower and higher couplings, associated with the transverse and longitudinal polarisations, respectively. This distinct behaviour of the different polarisation modes in the total luminosity ($L_{Z'}$) leads to noticeable differences in the resulting SN cooling constraints compared to existing studies~\cite{Croon:2020lrf, Lai:2024mse}.

    \item We incorporate the degeneracy effects of neutrinos in the calculation of production and absorption rates, rather than relying on vacuum rates as assumed in earlier studies~\cite{Croon:2020lrf, Lai:2024mse}. 
    
    \item We compute the attenuation factor numerically from absorption rates rather than using an approximation~\cite{Croon:2020lrf} and highlight the energy transport distribution in the $(r,\theta)$ directions inside the SN core, which may provide further insight into the underlying energy transport mechanism.
    
\end{itemize}

\section{Production of \texorpdfstring{$\boldsymbol{Z^\prime}$}{Z'} in Supernovae}\label{sec:Prod}

\begin{figure}[ht]
	\begin{tikzpicture}[scale=2]

	\def\R{1.5}        
    \def\rp{1.1}
	\def\r{0.5}        
	\def\angle{30}    
    \def\a2{25}
	\def\rnu{0.7}
    \def\exitx{{\R*cos(\angle)}}
    \def\exity{{\R*sin(\angle)}}
    \def\midx{{\rp*cos(\a2)}}
    \def\midy{{\rp*sin(\a2)}}

	\coordinate (O) at (0,0) node[below left] {$O$};

	\draw[thick] (O) circle (\R);
	\filldraw[black] (O) circle (0.03);
	\draw[thick,dashed] (O) circle (\rnu);
	\draw[dashed,thick,<->] ($(O)+(0,0.03)$) -- (0,\rnu)node[midway, left] {$R_\nu$};

	\draw[dotted] (-\R,0) -- (\R,0);

	\coordinate (P) at (\r,0);
	\filldraw[black] (P) circle (0.03);
	\node[below] at (P) {$P$};

	\draw[->] (O) -- ($(P)-(0.03,0)$)
	node[midway,below] {$r$};

	\draw[thick,->]
	(P) -- (\exitx,\exity)
    node[pos=0.8, below, xshift=3pt] {$r_\alpha$}
	node[near end, xshift=18pt,yshift=14pt] {$E$};
    
    \draw[->]
	(O) -- (\midx,\midy)
    node[pos=0.4, above] {$r''$};

    \draw[->]
	(P) -- (\midx,\midy)
    node[pos=0.6, below] {$r'$};
    
	\draw (P) ++(0.4,0)
	arc[start angle=0, end angle=\angle, radius=0.6];
	\node at ($(P)+(0.5,0.2)$) {$\alpha$};

	\draw[<->] ($(O)+(0.03,-0.03)$) -- ({\R*cos(-45)},{\R*sin(-45})
	node[midway,sloped, below, xshift=15pt] {$R_{\rm far}$};

\end{tikzpicture}
	\caption{Diagram showing the neutrinosphere (dashed circle) with radius $R_\nu$ and the sphere of energy loss (solid circle) with $R_{\rm far}$. The LGB is produced at $P$ at a distance $r$ from the centre of the core and traverses a length $r_\theta$ along an angle $\theta$ from the radial direction and exits the sphere of energy loss at $E$. The variables $r'$ and $r''$ denote the integration parameters entering the attenuation factor in Eq.\eqref{eq:LumPol}, where $r'$ parametrises the trajectory along $\overrightarrow{PE}$ and $r''$ represents the radial distance of the infinitesimal element $dr'$ from the core centre.}
	\label{fig:snegeometry}
\end{figure}
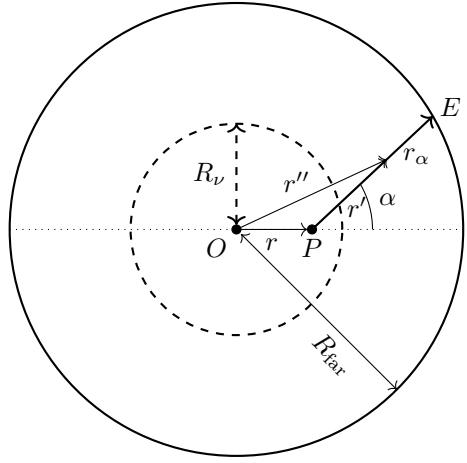

As already mentioned, in the absence of mixing,  one should calculate the luminosities for each polarisation independently. The same can be evaluated as~\cite{Chang:2016ntp,Croon:2020lrf,Lai:2024mse},
\begin{align}
	L_{Z'}^\lambda (g',M_{Z'}) =\int\limits_{\rm core} dV\int& \frac{d^3p}{(2\pi)^3}\omega~\Gamma_{\rm prod}^{\lambda}(\omega,r;~g',M_{Z'})\nonumber\\
    \times&\left[\frac{1}{2}\int\limits_{-1}^1d\cos\alpha~~{\exp}\left(-\int\limits_0^{r_\alpha}\frac{dr'}{|v|}\Gamma_{\rm abs}^{\lambda}(\omega,r''(r,r',\alpha);~g',M_{Z'})\right)\right] 
	\label{eq:LumPol}
\end{align}
where $\Gamma_{\rm prod}^\lambda(\Gamma^\lambda_{\rm abs})$ is the production (absorption) rate of the DP with $\lambda$ denoting the LGB polarisation ($\lambda \in \{L, T\}$) and $|v|=\sqrt{1-(M_{Z'}/\omega)^2}$ is the speed of the LGB. $r''(r,r',\alpha)$ is the distance of the line element $dr'$ along the LGB trajectory from the center of the core and $r_{\alpha}$ is the distance between the production location and the surface of the sphere with radius $R_{\rm far}$ along the direction $\alpha$ from radially outward direction as shown in Fig.\,\ref{fig:snegeometry}. 
Here, $R_{\rm far}$ denotes the radius beyond which the energy carried by the LGB can no longer be reprocessed into neutrinos and is therefore considered lost from the PNS. The term in the square brackets, referred to as the \emph{attenuation factor}, represents the mean probability for produced LGBs to escape the $R_{\rm far}$-sphere averaged over all propagation directions ($\cos\alpha$). It accounts for losses due to reabsorption in the core as well as decays occurring before the LGBs exit the sphere. When the attenuation factor is unity, the LGB is free-streaming, whereas a vanishing attenuation factor indicates that the produced LGB is entirely trapped within the medium. The averaging over directions makes the volume integrand spherically symmetric, so that one may write, $\int dV\to\int4\pi r^2 dr$.

\subsubsection*{The Trapping Ragime}
Once the coupling is strong enough, the LGB production and absorption become comparable. This results in the LGB equilibrating with the SM bath in the inner core and, analogous to neutrino-sphere, forms a \emph{LGB-sphere}. At this coupling, the LGB distribution $(f^\lambda)$ cannot be neglected and stimulates further LGB emissions~\cite{Caputo:2021rux}. The energy transport due to LGBs reads:
\begin{align}
	\partial_tf^\lambda+\partial_sf^\lambda
	 & =\Gamma^\lambda_{\rm prod}(1+f^\lambda)-\Gamma_{\rm abs}f^\lambda\nonumber                        \\
	 & =\Gamma_{\rm prod}^\lambda-(\Gamma_{\rm abs}^\lambda-\Gamma_{\rm prod}^\lambda)f^\lambda\nonumber \\
	 & =\Gamma_{\rm prod}^\lambda-\Gamma_{\rm abs}^{\lambda, \rm eff}f^\lambda~~.
\end{align}
Since detailed balance implies that within the region of SM equilibrium, $\Gamma_{\rm prod}^\lambda=e^{-\omega/T}~\Gamma_{\rm abs}^\lambda$, the effective absorption rate ($\Gamma_{\rm abs}^{\lambda,\rm eff}$), entering the attenuation factor in Eq.\eqref{eq:LumPol}, can be evaluated as,
\begin{align}
	\Gamma_{\rm abs}^{\lambda,\rm eff}=\begin{cases}
	    (1-e^{-\omega/T})\Gamma_{\rm abs}^\lambda~,& r\le R_\nu\\
        \Gamma_{\rm abs}^\lambda~,& r>R_\nu ~~
	\end{cases}
\end{align}


\section{The Gauged \texorpdfstring{$L_\mu - L_\tau$}{L\_mu - L\_tau} Model}\label{sec:DPModel}
In this section, we present the $U(1)_{L_\mu - L_\tau}$ model and compute the production and absorption rates of light $Z^\prime$ in the supernova environment. The $2$nd generation ($Q_\mu=+1$) and $3$rd-generation ($Q_\tau=-1$) doublet and singlet leptons of the SM are charged under this local abelian gauge symmetry~\cite{He:1990pn, Ma:2001md, Baek:2001kca}. All other SM fields are neutral under the symmetry. An additional scalar singlet ($\phi$) carrying the $L_\mu-L_\tau$ charge ($Q_\phi$) is required to generate the $Z^\prime$ mass through its vacuum expectation value after the $U(1)_{L_\mu - L_\tau}$ breaking (\emph{i.e.}, $M_{Z^\prime}= Q_\phi ~g^\prime v_\phi$).  The anomaly-free scenarios for a typical $U(1)_X$ model require additional chiral fermions. In contrast, the $L_\mu-L_\tau$ model does not require any right-handed neutrino for anomaly cancellations. The interaction Lagrangian relevant to our discussion after symmetry breaking is given by:
\begin{align}
\mathcal{L}_{Z^\prime}^{\rm int.} &\supset g^\prime Z^\prime_\mu J_{L_\mu-L_\tau}^{\mu} + (\epsilon e) Z^\prime_\mu J_{\rm EM}^{\mu} \nonumber \\
&= g^\prime Z^\prime_\mu \Big( \bar \mu \gamma^\mu \mu + \bar \nu_\mu \gamma^\mu P_L \nu_\mu - \bar \tau \gamma^\mu \tau - \bar \nu_\tau \gamma^\mu P_L \nu_\tau \Big)+(\epsilon e) Z^\prime_\mu \sum_{f \in {\rm SM}}  Q_f^{\rm EM} ~\bar f \gamma^\mu f ~~~ .
\end{align}
where $f$ stands for the charged fermions in the SM. The first term corresponds to the tree-level interaction with $Z^\prime$ present in the model, while the second term arises from the induced coupling $\epsilon$. The source of the induced coupling depends on the underlying model. We adopt $\epsilon \approx -{g^\prime}/{70}$ for our discussion, which is similar to a typical loop-induced interaction for $\mdp<2m_\mu$~\cite{Escudero:2019gzq}.

 These gauge bosons are among the simplest and well-motivated extensions of the SM, as they can be probed across a broad mass range spanning from eV to TeV\footnote{The development of grand unified theories (GUTs) that go beyond the original $SU(5)$ model, such as those based on $SO(10)$ or $E_6$ provides strong motivation for these extensions \cite{Hewett:1988xc,Langacker:2008yv}.}. Collider searches (LHC, LEP) dominate at high masses, while LHCb, Belle, BaBar, fixed-target experiments, and late time cosmological observations constrain the MeV–GeV range \cite{ATLAS:2019erb,Essig:2009nc,LHCb:2017trq,BaBar:2014zli,Bross:1989mp,Escudero:2019gzq,Ghosh:2024cxi}. In this work, we focus on the sub-GeV mass region (eV to $\sim 100$ MeV), where the production of LGBs inside the supernova core is relevant. Below, we discuss in detail the production of LGBs inside the supernova core.

\subsection{Production and Absorption rates}\label{sec:Rates}
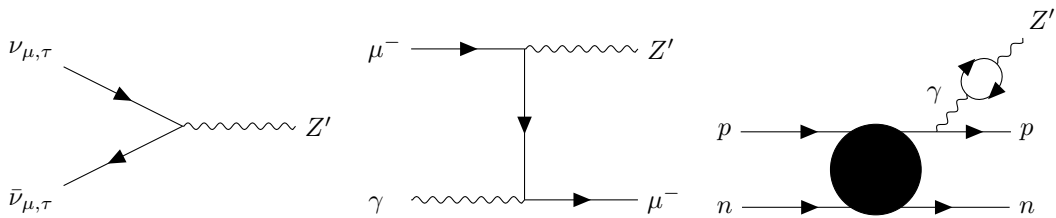
\begin{figure}[ht]
	\centering
		\begin{tikzpicture}
		\begin{feynman}
			\vertex (v) at (0,0);
			\vertex (e1) at (-2, 1) {\(\nu_{\mu,\tau}\)};
			\vertex (e2) at (-2,-1) {\(\bar\nu_{\mu,\tau}\)};
			\vertex (Ap) at ( 1.8, 0) {\(Z'\)};

			\diagram* {
			(e1) -- [fermion] (v),
			(e2) -- [anti fermion] (v),
			(v)  -- [photon] (Ap),
			};
		\end{feynman}
	\end{tikzpicture}~
	\begin{tikzpicture}
		\begin{feynman}
			\vertex (v);

			\vertex [below=2cm of v] (u);
			\vertex [left=of v]  (ei) {\(\mu^-\)};
			\vertex [right=of v] (go) {\(Z'\)};
			\vertex [right=of u]  (ef) {\(\mu^-\)};
			\vertex [left=of u] (gi) {\(\gamma\phantom{^-}\)};

			\diagram* {
			(ei) -- [fermion] (v) -- [fermion] (u) -- [fermion] (ef),
			(v) -- [photon]  (go),
			(gi)  -- [photon] (u),
			};
		\end{feynman}
	\end{tikzpicture}~
	\begin{tikzpicture}
		\begin{feynman}

			\vertex (n_in) at (-2, -0.5) {$n$};
			\vertex (p_in) at (-2, 0.5) {$p$};

			\vertex (n_mid) at (0, -0.5);
			\vertex (p_mid) at (0, 0.5);

			\vertex (g_start) at (0.8, 0.5);
			\vertex[blob, minimum size=1.2cm, black] (blob) at (0,0) {};

			\vertex (n_out) at (2,  -0.5) {$n$};
			\vertex (p_out) at (2, 0.5) {$p$};

			\vertex (lstrt) at (1.2,1);
			\vertex (lend) at (1.6,1.4);

			\vertex (gamma) at (2.2, 2) {$Z'$};

			\diagram*{
			(n_in) -- [fermion] (n_mid) -- [fermion] (n_out),
			(p_in) -- [fermion] (p_mid) --  (g_start) -- [fermion] (p_out),
			(g_start) -- [photon, edge label={$\gamma$}] (lstrt) -- [fermion, half left, looseness=1.5] (lend) -- [fermion, half left, looseness=1.5] (lstrt),
			(lend) -- [photon] (gamma),
			};

		\end{feynman}
	\end{tikzpicture}
	\caption{Dominant dark photon production channels inside the core of the supernova via pair coalescence (left), semi-Compton scattering ($s$-channel diagram not shown) (middle), and loop Bremstrahlung of neutron-proton scattering (right).}
	\label{fig:Feyn2}
\end{figure}

Within the core, the LGB is dominantly produced via three channels as shown in Fig.\,\ref{fig:Feyn2}, {\it viz.} pair-coalascence  (left), semi-Compton (middle), and loop-bremstrahlung (right). The LGB can be reabsorbed within the core through reverse processes of these channels. In the following, we discuss and derive the rates for each of these processes.

\subsubsection{Pair-coalescence}
In earlier studies, the rate of the pair-coalescence (PC) and inverse PC (decay) process is generally evaluated assuming non-degenerate neutrinos~\cite{Croon:2020lrf,Lai:2024mse}. However, within the neutrinosphere, muon neutrinos are highly degenerate. In the {\tt SFHo-18.8} model~\cite{Steiner:2012rk,Bollig:2020xdr}, the muon-neutrino chemical potential approaches 50 MeV near the center\footnote{We use data extracted from the plots provided in Refs.~\cite{Bollig:2020xdr,Caputo:2021rux} for our calculations. The data is available at \href{https://wwwmpa.mpa-garching.mpg.de/ccsnarchive/}{https://wwwmpa.mpa-garching.mpg.de/ccsnarchive/}.}. Tau neutrinos, on the other hand, can be treated as non-degenerate, with negligible chemical potential.
Therefore, in this work, we reevaluate PC production and decay rates by consistently incorporating degeneracy effects. The production rate for the PC process is given by (see Appendix-\ref{app:phasespace}),
\begin{align}
	\Gamma_{\rm prod}^{\lambda,\,\rm PC}
	 & =\sum_{\nu_\mu,\nu_\tau}\frac{1}{16\pi\omega}\int d\cos\theta_\nu \,f_{\nu}\left(E_\nu\right)f_{\bar\nu}\left(\omega-E_\nu\right)\frac{2E^2_{\nu}}{M^2_{Z'}}|\mathcal M_{\nu\bar{\nu}\to Z'_\lambda}|^2\Bigg|_{E_\nu=\frac12{M_{Z'}}/{(\cosh \alpha \,-\,\cos \theta_\nu \,\sinh \alpha  )}}
\end{align}
where $\alpha$ is the rapidity of LGB in SN plasma frame, \emph{i.e.} $\cosh\alpha=\omega/M_{Z^\prime}$, $\theta_\nu$ is the angle at which the neutrino is emitted relative to the direction of LGB propagation, $f_{\nu,\bar\nu}$ are the equilibrium thermal distribution of neutrinos and antineutrinos in the core.  $\mathcal M_{\nu\bar{\nu}\to Z'_\lambda}$ is the scattering matrix element corresponding to LGB polarisation $\lambda$  
and are given by,
\begin{align}
	\left|\mathcal M^{L,PC}\right|^2   & ={g^\prime}^2 M_{Z'}^2\left[\frac{  \sin\theta_\nu  }{\cosh\alpha-\sinh \alpha ~ \cos \theta_\nu  }\right]^2,                         \\
	\left|\mathcal M^{T_i,PC}\right|^2 & =\frac12 {g^\prime}^2 M_{Z'}^2 \left[1+\left(\frac{ \cos \theta_\nu  -\tanh \alpha }{1-\tanh \alpha ~ \cos \theta_\nu}  \right)^2\right]~~.
	\label{eq:PCrate}
\end{align}

The rate for the inverse of pair-coalescence, {\it i.e.}, LGB decay to $\nu+\bar\nu$, can be obtained by employing the principle of detailed balance.
Outside the neutrinosphere, assuming $f_{\nu,\bar\nu}(r>R_\nu)\approx0$, the finite temperature effect can be neglected. The LGB can thus only be annihilated through decay, the rate for which is essentially the vacuum decay rate. Since neutrinos in this region are not in thermal equilibrium, the principle of detailed balance does not hold, and the production rate may be neglected. The decay rate can thus be approximated as,
\begin{align}
	\Gamma_{\rm abs}^{\lambda,\rm PC}=
	\begin{cases}
		e^{\omega/T}~\Gamma_{\rm prod}^{\lambda,\rm PC},
		                                   & r\leq R_\nu \\[7pt]
		\dfrac{g'^2M^2_{Z'}}{12\pi\omega}, & r>R_\nu
	\end{cases}\label{eq:decay}
\end{align}
\begin{figure}[htb!]
    \centering
    \includegraphics[width=0.49\linewidth]{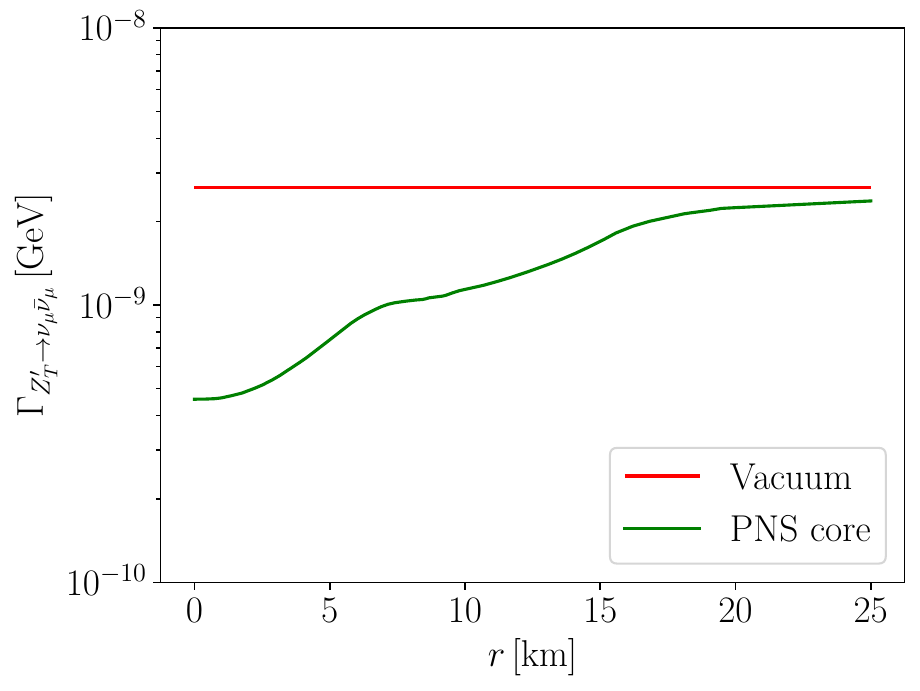}
    \includegraphics[width=0.49\linewidth]{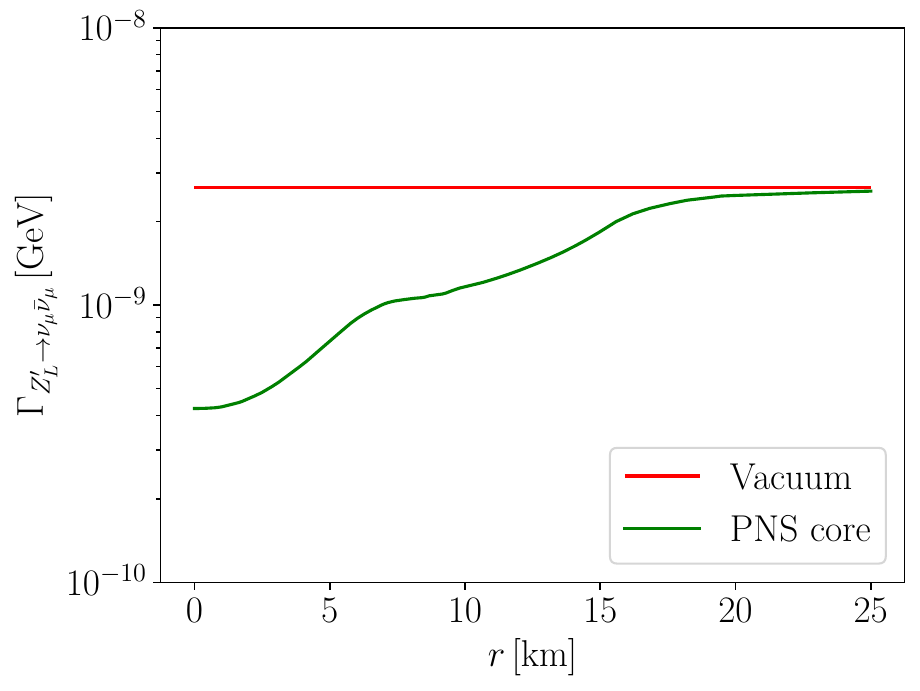}
    \caption{Comparison of the LGB decay rates obtained using thermal $\nu_\mu$ and $\bar\nu_\mu$ distributions in the PNS core, assuming {\tt SFHo-18.8} model~\cite{Steiner:2012rk, Bollig:2020xdr} for SN1987A, against the vacuum decay rate of the LGB for $T$-modes (left) and $L$-mode (right) for $\omega=50$\,MeV, $\mdp=100$\,keV and $g'=1$.}
    \label{fig:diff_vac_SN_PC}
\end{figure}

To illustrate the above, Fig.\,\ref{fig:diff_vac_SN_PC} shows the slight difference between the thermal decay rate of LGB to $\nu_\mu\bar\nu_\mu$ from the vacuum decay rate for $T$ (left) and $L$ (right) modes of the LGB for $\omega=50$\,MeV, $\mdp=100$\,keV and $g'=1$ as nominal values. The chemical potential for neutrinos is taken from the {\tt SFHo-18.8} model~\cite{Steiner:2012rk, Bollig:2020xdr} of the SN1987A. As is clear from the figure, the Pauli-blocking due to degeneracy in the medium starts to disappear, and the decay rate approaches the vacuum decay rate in the outer region where the neutrinos are non-degenerate. {The PC process of muons is kinematically forbidden for $\mdp<2m_\mu$}.

\subsubsection{Semi-Compton}
We first derive the expressions for the absorption rates of the LGB polarisations through the semi-Compton (SC) process and then employ the principle of detailed balance to evaluate the production rate. This absorption rate is expressed as~\cite{Croon:2020lrf},
\begin{align}
	\Gamma_{\rm abs}^{\lambda, \rm SC}=\left(\frac{1}{1-e^{-\omega/T}}\right)\times n_\mu F_{\rm deg}\times\sqrt{1-\frac{M_{Z'}^2}{\omega^2}}~\sigma^{\lambda}_{\rm SC}~.	\label{eq:SCrate}
\end{align}
The cross-section for individual polarisation in the limit $\mdp\ll\omega\ll m_\mu$ is given as\footnote{Since the SC process is only dominant for small LGB mass, this approximation remains valid. For $M_{Z'}\sim m_\mu$, PC/decay dominates the total rate.}:
\begin{align}
	\sigma^{T_i,L}_{\rm SC} & =\frac{8\pi\alpha\alpha'}{3m_\mu^2}\times\left[\frac{M_{Z'}^2}{\omega^2}\right]_L ~,
\end{align}
where $\alpha=e^2/4\pi$ and $\alpha'=g'^2/4\pi$. The term in the square bracket is multiplied only for the longitudinal mode. The production rate can be obtained using the principle of detailed balance as,
\begin{align}
	\Gamma_{\rm prod}^{\lambda, \rm SC} & =e^{-\omega/T}~\Gamma_{\rm abs}^{\lambda, \rm SC}\nonumber                                                                   \\
	                                   & =\left(\frac{1}{e^{\omega/T}-1}\right)\times n_\mu F_{\rm deg}\times\sqrt{1-\frac{M_{Z'}^2}{\omega^2}}~\sigma^{\lambda}_{\rm SC}~.
\end{align}
For numerical analysis, we adopt the values of muon degeneracy factor $F_{\rm deg}$, muon number density $n_\mu$, and temperature $T$ from Ref.~\cite{Bollig:2020xdr}. 

\subsubsection{Loop-bremsstrahlung}
In the $L_\mu-L_\tau$ model, the LGB does not interact with the proton at tree level. However, the LGB can couple to electromagnetic currents through an induced coupling ($\epsilon$), specifically via a muon-induced loop in our discussion. The loop suppression can be overcome by the large densities of neutrons ($n_n$) and protons ($n_p$) available in the core. In the soft radiation approximation limit, the width for the inverse loop bremsstrahlung (LB) process is \cite{An:2013yfc,Rrapaj:2015wgs,Chang:2016ntp},
\begin{align}
	\Gamma_{\rm prod/abs}^{\lambda,\rm LB} & =\frac{32}{3\pi}\frac{\alpha\epsilon^2n_nn_p}{\omega^3}\left(\frac{\pi T}{m_N}\right)^\frac32\langle\sigma^{(2)}_{np}(T)\rangle_{\rm prod/abs}\times\left[\frac{M_{Z'}^2}{\omega^2}\right]_L,\label{eq:LBrate}
\end{align}
where the term in the square bracket is multiplied only for the longitudinal polarisation. In the  above expression, $\alpha$ is the electromagnetic fine structure constant, $\epsilon$ is the mixing of SM photon and LGB related to the coupling at one-loop as $\epsilon\approx -g'/70$ for $M_{Z'}\ll m_\mu$~\cite{Escudero:2019gzq} in absence of an explicit mixing term in the interacting Lagrangian\footnote{Medium effects on SM photons are expected to modify the mixing parameter $\epsilon$~\cite{An:2013yfc, Chang:2016ntp}, but are not considered in this work.}, and
\begin{align}
	\langle\sigma^{(2)}_{np}(T)\rangle_{\rm abs}=\frac12\int_{-1}^1d\cos\theta_{np}\int_0^\infty dx~x^2e^{-x}(1-\cos\theta_{np})\frac{d\sigma_{np}(xT)}{d\cos\theta_{np}}
\end{align}
is the avearged $np$ dipole scattering cross-section. The differential cross-section for the $np$-scattering is calculated using the {\tt reid93} potential~\cite{Zhaba:2018iyj} and the results are obtained from the webpage~\cite{nnonline}. For the production rate, the center of mass energy is required to be $T_{\rm CM}=x ~T>\omega$, modifying the lower limit on the integral,
\begin{align}
	\langle\sigma^{(2)}_{np}(T)\rangle_{\rm prod}=\frac12\int_{-1}^1d\cos\theta_{np}\int_{\omega/T}^\infty dx~x^2e^{-x}(1-\cos\theta_{np})\frac{d\sigma_{np}(xT)}{d\cos\theta_{np}}
\end{align}
For a detailed discussion on the bremsstrahlung process in SNe, the reader is referred to refs.~\cite{An:2013yfc, Chang:2016ntp, Rrapaj:2015wgs}.
\section{Results and Discussions}\label{sec:Res}
In this section, we present and discuss the results for the luminosity ($L_{Z'}$) due to the LGB with mass $\mdp$ and coupling $g'$ in the SN environment. The $L_{Z'}^\lambda$ is numerically estimated independently for each polarisation mode $\lambda$, following Eq.\eqref{eq:LumPol},  based on the above discussion of the production and absorption rates, 
\begin{align}\Gamma^\lambda_{{\rm prod}/{\rm abs}}=\Gamma^{\lambda,{\rm PC}}_{{\rm prod}/{\rm abs}}+ \Gamma^{\lambda,{\rm SC}}_{{\rm prod}/{\rm abs}}+ \Gamma^{\lambda,{\rm LB}}_{{\rm prod}/{\rm abs}}~,\hspace{1cm}\lambda \in \{T_{i=1,2},L\}.
\end{align}
We focus on regions with $g'<0.1$, as a larger coupling would lead to additional processes, such as \emph{dark-Compton} scattering $\mu+Z'\leftrightarrow\mu+Z'$, allowing mixing between the $L$- and $T$-modes. We set $R_{\rm far}=100$\,km as a reference value and later explore other values. 
To evaluate the luminosity numerically, we first compute the integral appearing in the attenuation factor of Eq.\eqref{eq:LumPol},
\begin{align}
I^\lambda(\omega,r,\alpha,g',\mdp)
&=
\int_0^{r_\alpha}
\frac{dr'}{|v|}
\Gamma_{\rm abs}^\lambda
\left(
\omega,
r''(r,r',\alpha),g',\mdp
\right)
\nonumber\\
&=
g'^2
\int_0^{r_\alpha}
\frac{dr'}{|v|}
\Gamma_{\rm abs}^\lambda
\left(
\omega,
r''(r,r',\alpha),
1,\mdp
\right)
\nonumber\\
&\equiv
g'^2\times I_2^\lambda(\omega,r,\alpha,\mdp),
\end{align}
where we have used the fact that the absorption rate scales as
$\Gamma_{\rm abs}^\lambda\propto g'^2$.
Assuming \(I_2^\lambda\) to be a smooth function of its arguments, we precompute it on a multidimensional grid in
\((\omega,r,\alpha,\mdp)\) of size $100^4$ for each polarization mode.
Values between grid points are obtained using multidimensional interpolation implemented through the {\tt RegularGridInterpolator} routine in the {\tt SciPy} package. Denoting the resulting interpolated function by \(\widetilde I_2^\lambda\), the luminosity can be written as
\begin{align}
    L_{Z'}^\lambda
    =
    \int dV
    \int\frac{d^3p}{(2\pi)^3}
    \int d\cos\alpha~~
    \omega\,\Gamma_{\rm prod}^\lambda(\omega,r,g',\mdp)\times\frac{1}{2}
    \exp\!\left[
    -g'^2\,
    \widetilde I_2^\lambda
    (\omega,r,\alpha,\mdp)
    \right].
\end{align}
The remaining multidimensional integral is evaluated using a Monte Carlo (MC) integration technique. 

\begin{figure}[h!]
	\centering
	\includegraphics[width=0.8\linewidth]{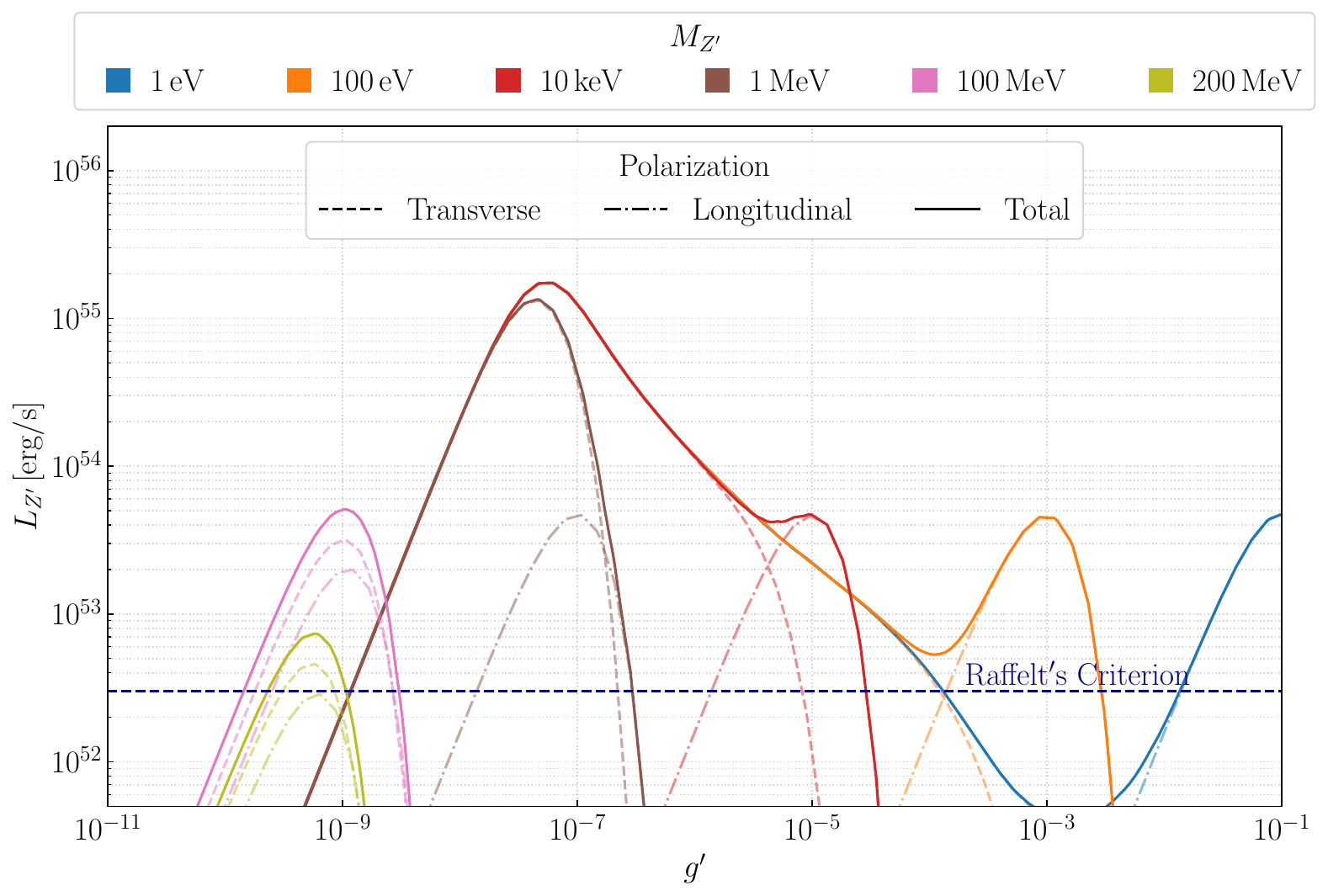}
	\caption{The luminosity of the LGB as a function of the gauge coupling $g^{\prime}$ for different $M_{Z'}$ values is shown in different colours. The horizontal blue dotted line indicates the Raffelt limit ($L_{Z'} \simeq 3 \times 10^{52}$ erg/s), with the region above excluded based on observations from SN1987A. The dashed and dot-dashed line represents the contribution of transverse ($2 L_{Z'}^T$) and longitudinal ($L_{Z'}^L$) polarisations of $Z'$, respectively. The solid line represents the total luminosity, $2 L_{Z'}^T+ L_{Z'}^L$~.}
	\label{fig:lmutauExcl}
\end{figure}

Fig.\,\ref{fig:lmutauExcl} shows the expected luminosity of LGB for six distinct benchmarks (BMPs) of $\mdp$ as listed in Table-\ref{tab:BMPs}, from the SNe, assuming the SFHo-18.8 model, at 1\,sec post-bounce, corresponding to the epoch when the temperature typically reaches its peak \cite{Steiner:2012rk,Bollig:2020xdr}. 
The solid lines show the total luminosity ($L_{Z'}=2 L_{Z'}^T + L_{Z'}^L $), while the dashed (dot-dashed) lines show the contribution from the two transverse (one longitudinal) polarisations. The colours represent the corresponding values of $\mdp$.
The blue dashed horizontal line denotes the maximum allowed luminosity according to Raffelt's criterion $(L_{Z'} = 3\times10^{52}\,\rm erg/s)$ \cite{Raffelt:1996wa}. For a fixed $\mdp$, the coupling strength that yields a luminosity exceeding this limit is excluded.
\begin{table}[h!]
    \centering
    \begin{tabular}{c ccccccc}
        \toprule
        BMP & 1 & 2 & 3 & 4 & 5 & 6 \\
        \midrule
        $\mdp$ ~& ~200\,\text{MeV} ~&~ 100\,\text{MeV} ~&~ 1\,\text{MeV} ~&~ 10\,\text{keV} ~&~ 100\,\text{eV} ~&~ 1\,\text{eV} \\
        \bottomrule
    \end{tabular}
    \caption{Benchmark masses of LGB for luminosity estimation.}
    \label{tab:BMPs}
\end{table}

As evident from Fig.\,\ref{fig:lmutauExcl}, the LGB luminosity $L_{Z'}^{\lambda}$ increases at first with the coupling strength $g'$, due to the enhanced production rate.
This region of smaller coupling for a given $\mdp$ is called the \emph{free-streaming} region due to the minimal reabsorption of $Z'$. As the coupling strength increases further, the absorption rate starts to affect the number of LGBs that escape the $R_{\rm far}$-sphere.
Consequently, the luminosity starts to reduce and dies exponentially. This region is called the {\it trapping} regime, as the LGB is effectively trapped within the SNe environment.

 The Fig.\,\ref{fig:lmutauExcl} shows different luminosity characteristics depending on the LGB masses. BMPs {\bf 1} and {\bf 2} exhibit noticeably smaller luminosity peaks compared to the other benchmarks. This is because, for relatively large masses, the production and absorption rates are dominated by the PC and inverse PC (decay) processes, respectively, with $\Gamma^{\rm PC} \propto g'^2 \mdp^2$ (see Eqs.~\eqref{eq:PCrate} and \eqref{eq:decay}). As $\mdp$ exceeds the typical thermal energy scale of the plasma ($\sim 30\,\mathrm{MeV}$), the produced LGBs are non-relativistic and carry low kinetic energies. Consequently, they spend a longer time within the $R_{\rm far}$-sphere, reducing their mean free path ($\lambda_{\rm mfp} \propto |v|/\Gamma_{\rm abs}$) and, equivalently, their probability of escaping the medium. This leads to suppressed luminosity peaks for all polarisation modes.
Additionally, although $|\mathcal{M}|^2$ is larger for BMP {\bf 1} than for BMP {\bf 2} due to the larger value of $\mdp$, the resulting luminosity is smaller even in the \emph{free-streaming} regime. This can be attributed to the stronger Boltzmann suppression in the production of the heavier LGB in BMP {\bf 1}.

In contrast to BMPs {\bf 1} and {\bf 2}, the luminosity of the $T$-modes in BMPs {\bf 3}--{\bf 6} is identical in the free streaming regime (smaller couplings) and only differs in the trapping regime (larger couplings). 
The luminosity becomes independent of mass in the \emph{free-streaming regime} because the production rates of the $T$-modes (dashed lines) are dominated by the SC process, whose rate does not depend on $M_{Z'}$.
Unlike the \emph{free-streaming} case, the \emph{trapping regime} shows mass dependence due to differences in absorption rates within the $R_{\rm far}$-sphere. This is due to the dependence of inverse SC on the muon density ($n_{\mu}$) in the environment (Eq.\eqref{eq:SCrate}).
While inverse SC absorption is confined to the muon-rich core ($r\lesssim 20\,\mathrm{km}$, as shown later), reabsorption in the outer, muon-poor region ($20\,\mathrm{km}\lesssim r < R_{\rm far}$), which spans a much larger volume, is dominated by decay channels (inverse PC). Since the decay rate grows with mass for a given coupling (Eq.\eqref{eq:decay}), the LGBs with larger $\mdp$ start to get trapped for relatively smaller couplings, reducing the luminosities, as shown in the figure.
Additionally, this limited spatial extent of inverse SC reabsorption ($r\lesssim20$\,km), together with a small decay rate compared to production (through SC), results in higher peaks (for BMPs {\bf 3}--{\bf 6}) as compared to the PC-dominated production in BMPs 1 and 2. The loop-bremsstrahlung (LB) process is subdominant in the entire parameter space due to its suppressed induced mixing $\epsilon$, despite the large nucleon densities in the SN environment\footnote{This is consistent with the results of Ref.~\cite{Croon:2020lrf}, where only PC and SC processes were considered, while for LB, complementary bounds were adopted from Ref.~\cite{Chang:2016ntp} after rescaling.  
}. Nevertheless, it aids the SC reabsorption for $T$-modes within the core. As a result, the luminosity falls more rapidly compared to the case when it is neglected.

Unlike $T$-modes, the production and attenuation channels for the $L$-mode for all the BMPs are dominated by the PC and inverse-PC (decay) process ($\propto  \mdp^2/\omega$), as SC and LB channels are suppressed ($\propto \mdp^2/\omega^2$, see Eq.\eqref{eq:SCrate} and \eqref{eq:LBrate}). The dominant PC contribution shifts the luminosity curve towards a larger $g'$ at a lower mass, as the PC rate scales as $\Gamma^{\rm PC}\sim g'^2\mdp^2$.
In BMP {\bf 3}, although the peaks of the luminosity curves for both modes ($2 T$ $\&$ $L$) appear in a similar coupling range, the $2~T$-modes provide a significantly larger contribution to the total $Z'$ luminosity than the $L$-mode as the former is SC dominated while the latter is PC dominated. 

 A direct consequence of considering suppressed polarisation intermixing rates is evident in the total luminosity ($L_{Z'}=2 L_{Z'}^T +  L_{Z'}^L $) for BMPs with very low masses. In particular, in BMPs {\bf 4}, {\bf 5}, and {\bf 6}, the luminosity exhibits two distinct peaks. 
This can be attributed to the large differences in the rates of the $L$- and $T$-modes. At lower couplings, the $T$-modes are efficiently produced via the SC process, as discussed earlier, and dominate the emission ($2 L_{Z'}^T$). However, as the coupling increases, these modes become trapped due to their large absorption rates. In contrast, the production of $L$-modes is suppressed at low couplings because of small $\mdp$, and requires larger couplings to become significant.
Therefore, the luminosity at low couplings is dominated by the $T$-modes, whereas at higher couplings the $L$-mode contribution becomes dominant. 
As a result, two peaks appear in the total luminosity curve for lower $M_{Z'}$, with the second peak arising from a rise in luminosity at larger couplings due to the $L$-mode contribution. 
This key feature in the total luminosity, illustrated in Fig.\,\ref{fig:lmutauExcl} by the red, orange, and cyan solid lines for the respective LGB masses, significantly modifies the SN cooling constraints, which will be discussed later.

\begin{figure}
\begin{tikzpicture}[scale=0.1]

\draw[thick] (0,0) circle (20);
\draw[dotted] (0,0) circle (10);
\node at (0,-28) {$\nu$-Sphere};

\draw[->] (0,0) -- (20,0) node[above, pos=0.7]{$R_\nu$};

\coordinate (P) at ({14*cos(130)},{14*sin(130)});

\draw[thick, ->] (0,0) -- ($(P)-(-1,1)$) node[above, pos=0.4]{$r$};
\draw[dashed] ($(P)+(-1,1)$) -- ({21*cos(130)},{21*sin(130)}) ;
\draw[->,line width=0.5pt]
      ({2},{0})
      arc[start angle=0,end angle=130,radius=2] node[above, pos=0.4]{$\theta$};

\draw[->,line width=0.5pt]
      ({17*cos(130)},{17*sin(130)})
      arc[start angle=130,end angle=0,radius=3] node[above, pos=0.4]{$\alpha$};

\draw[rotate around={130:(P)}, thick] ($(P)+(-1,-1)$) rectangle ($(P)+(1,1)$);
\node[below left] at (P) {$dV$};

\draw[dashed,thick,->] ($(P)+(1,0)$) -- (99,0)  node[above, midway, sloped]{$r_\alpha(r,\alpha(r,\theta))$};



\node[above, right] at ({20*cos(0)},{20*sin(0)}) {0$^\circ$};
\node[above, above right] at ({20*cos(45)},{20*sin(45}) {45$^\circ$};
\node[above, above] at ({20*cos(90)},{20*sin(90}) {90$^\circ$};
\node[above, above left] at ({20*cos(135)},{20*sin(135}) {135$^\circ$};
\node[above, left] at ({20*cos(180)},{20*sin(180}) {180$^\circ$};
\node[above, below left] at ({20*cos(45*5)},{20*sin(45*5}) {225$^\circ$};
\node[above, below] at ({20*cos(45*6)},{20*sin(45*6}) {270$^\circ$};
\node[above, below right] at ({20*cos(45*7)},{20*sin(45*7}) {315$^\circ$};


\draw[line width=2pt]
      ({100*cos(3)},{100*sin(3)})
      arc[start angle=3,end angle=-3,radius=100];
\draw[dashed,line width=1pt]
      ({100*cos(10)},{100*sin(10)})
      arc[start angle=10,end angle=-10,radius=100];

\node[right] at (100,0) {$A_{\rm unit}=1\,\rm km^2$};
\node[right] at (99,14) {$R_{\rm far}$-sphere};


\draw[<->] (0,-33) -- (100,-33);
\node[below] at (50,-33) {$R_{\rm far}$};

\end{tikzpicture}
    \caption{Schematic illustration of the SN core (solid circle), the high-temperature shell (dotted circle) that dominates LGB production, and the $R_{\rm far}$-sphere. An arbitrary unit surface element, $A_{\rm unit}$, is chosen on the $R_{\rm far}$-sphere. The quantity $dL^\lambda_{Z'}/dV\,dA$ is evaluated to determine how different regions of the core contribute to the luminosity flux reaching $A_{\rm unit}$.}
    \label{fig:rnurfar}
\end{figure}
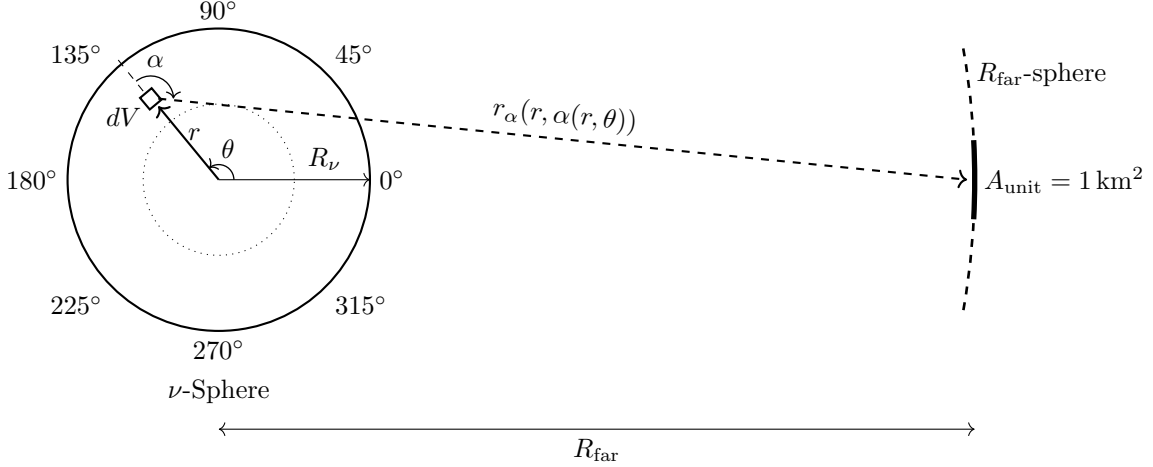

The hot shell located at $r\sim10$\,km inside the SN core is the primary production region for LGBs. After production, the LGBs propagate through regions with varying absorption rates before reaching the $R_{\rm far}$-sphere. The probability that an LGB reaches a given surface element on this sphere depends on its production location, the intervening medium, and the LGB parameters $(M_{Z'},g')$. Consequently, different regions of the core contribute differently to the luminosity flux through a given surface element.
To visualise this energy transport, we study the distribution of the contributions to the transverse-mode luminosity reaching a chosen unit surface area $A_{\rm unit}$ on the $R_{\rm far}$-sphere (see Fig.\,\ref{fig:rnurfar}). The luminosity expression in Eq.\eqref{eq:LumPol} can be rewritten as,
\begin{align}
    L_{Z'}^\lambda&=\int\limits_{\rm core} dV\int\limits_{M_{Z'}}^\infty d\omega~\frac{dL^\lambda_{Z'}}{dV\,d\omega}\\
    &=\int\limits_{R_{\rm far}} dA~\frac{d}{dA}\int\limits_{\rm core} dV~\frac{dL^\lambda_{Z'}}{dV} \nonumber \\
    &=4\pi R_{\rm far}^2\int\limits_{\rm core} dV~\left[\frac{dL^\lambda_{Z'}}{dV\,dA}\right]_{A_{\rm unit}}
    \label{eq:Lum2}
\end{align}
The quantity in square brackets represents the contribution to the escaping $\lambda$-mode luminosity through a unit surface area $A_{\rm unit}$, originating from a unit volume element located at $(r,\theta)$ in the core and is given by,
\begin{align}
    \frac{dL^\lambda_{Z'}}{dV\,dA}
    &=\frac{d\Omega}{dA}\frac{dL^\lambda_{Z'}}{dV\,d\Omega} \nonumber\\
    &=\frac{d\Omega}{dA}\int \frac{d^3p}{(2\pi)^3}
    \frac{\omega\,\Gamma_{\rm prod}^\lambda}{4\pi}
    \exp\!\left(-\int_0^{r_{\alpha}}
    dr'\frac{\Gamma_{\rm abs}^\lambda}{|v|}\right),\label{eq:dLdVdA}
\end{align}
where
\begin{align}
    \frac{d\Omega}{dA}
    =
    \frac{\left|r\cos\theta-R_{\rm far}\right|}
    {\left(r^2+R_{\rm far}^2
    -2rR_{\rm far}\cos\theta\right)^{3/2}}\label{eq:dOdA}
\end{align}
is the solid angle as seen from $(r,\theta)$, subtended by a unit surface element at the $R_{\rm far}$-sphere. We assume that the escape probability (the exponential factor) varies negligibly across the solid angle subtended by $A_{\rm unit}$. Without loss of generality, we choose $A_{\rm unit}$ to be the reference for polar coordinate $\theta$ (\emph{i.e.} $\theta_{A_{\rm unit}}\equiv0^\circ$, see Fig.\,\ref{fig:rnurfar}).
Note that, unlike Eq.\eqref{eq:LumPol}, where averaging over $\cos\alpha$ renders the volume integrand spherically symmetric ($dV=4\pi r^2dr$), the integrand in Eq.\eqref{eq:Lum2} corresponds to a specific pair of elements $(dV_{(r,\theta)},dA_{(R_{\rm far},0)})$, which fixes the propagation angle $\alpha=\alpha(r,\theta)$ (see Fig.\,\ref{fig:rnurfar}).
Consequently, the escape probability must be evaluated along that particular trajectory rather than averaged over directions. Additionally, the solid angle subtended by $A_{\rm unit}$ at $dV$ is also a function of $(r,\theta)$ (Eq.\eqref{eq:dOdA}).
The integrand $dL^\lambda/(dA\,dV)$ therefore depends on the polar angle $\theta$ and possesses only azimuthal symmetry, implying $\int dV\to\int2\pi r^2\,dr\,d\cos\theta$.

For illustration, we select three BMPs from Table-\ref{tab:BMPs} that show distinct behaviours in Fig.\,\ref{fig:lmutauExcl}, \emph{viz.} 
\begin{enumerate}
    \item \textbf{BMP 2}, which features a large LGB mass and is dominated by the PC production channel,
    \item \textbf{BMP 3}, in which the production of $T$-modes is dominated by the SC process, and,
    \item \textbf{BMP 5}, which is also SC-dominated but shows different trapping regime behaviour for $T$-modes compared to BMP 3, due to its smaller LGB mass.
\end{enumerate} 
For each BMP, we choose three distinct couplings around corresponding (i) \emph{free-streaming regime}, (ii) luminosity peak, and (iii) \emph{trapping regime} as listed in Table-\ref{tab:BMPcoup}.
\begin{table}[h!]
    \centering
    \begin{tabular}{ccccc}
    \hline
        BMP & $M_{Z'}$ & ~Free-streaming~ &~ ~Luminosity peak~ &~ Trapping regime~\\\hline
        BMP 2 & ~100\,MeV ~& $4\times10^{-10}$& $10^{-9}$& $3\times10^{-9}$ \\
        BMP 3 & 1\,MeV& $4\times10^{-9}$&$10^{-7}$& $2\times10^{-7}$ \\
        BMP 5 & 100\,keV&$4\times10^{-9}$&$10^{-7}$& $10^{-6}$\\\hline
    \end{tabular}
    \caption{Couplings in different regimes for each BMP to illustrate the $Z'$ parameter dependence of contributions to the luminosity from different regions of the core. These contributions are shown in Fig.\,\ref{fig:dLum}.}
    \label{tab:BMPcoup}
\end{table}

\begin{figure}[h!]
	\centering
	\includegraphics[width=0.9\linewidth]{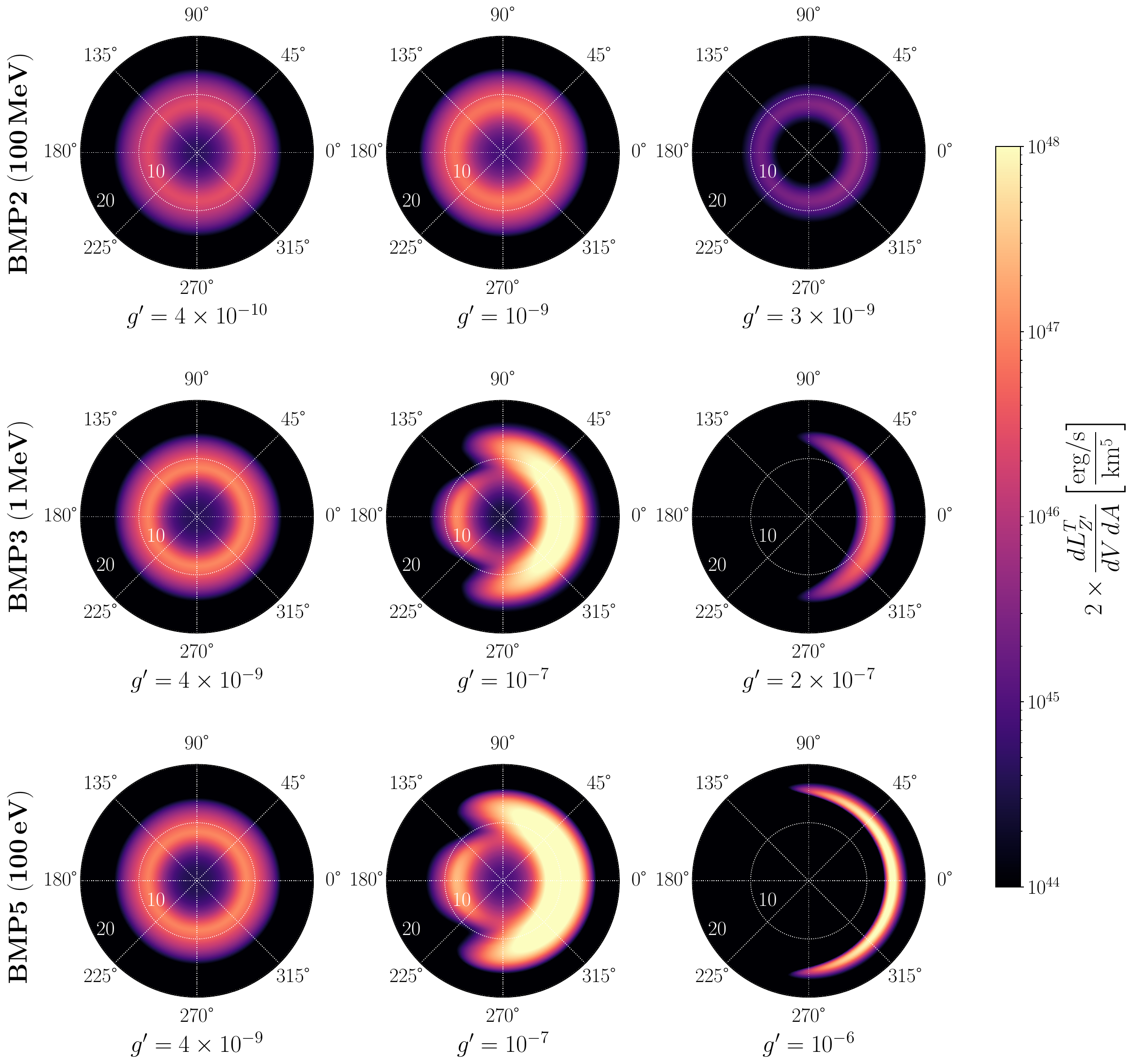}
	\caption{Contributions to $2T$-mode luminosity of LGBs in BMPs 2, 3, and 5, reaching $A_{\rm unit}$ (at $R_{\rm far}=100$\,km), from different regions of the core. The columns correspond to three different couplings specific to each BMP as listed in Table-\ref{tab:BMPcoup}, and are labelled below each plot. The colour bar on the right shows the numerical values, with bright indicating high contributions and dark indicating low contributions (black indicates regions with $2\,dL_{Z'}^T/dA\,dV<10^{44}\,\rm erg/s/km^5$).}
	\label{fig:dLum}
\end{figure}

Fig.\,\ref{fig:dLum} displays the heatmap of the above-mentioned contributions to luminosity flux for transverse modes of BMPs {\bf 2}, {\bf 3}, and {\bf 5}, from different locations in the core relative to $A_{\rm unit}$. Bright (yellow) regions correspond to larger LGB production with minimal attenuation, whereas dim (blue–black) regions either produce few LGBs or suffer substantial reabsorption before reaching the surface.
For small couplings (left column), the high temperature shell (at $\sim10$\,km) starts to glow of $Z'$, and the produced LGBs escape essentially unattenuated. 
At larger couplings (middle and right columns) the behaviour starts to deviate for the BMPs. In the case of BMP 2, the high-temperature shell first glows brighter (top-middle) and then gets dimmer as the coupling is increased further (top-right). 
This occurs because for this $\mdp$, the attenuation occurs primarily through LGB decay, with a rate that increases with coupling, as explained earlier. 
As the core medium only slightly reduces the decay rate (see Fig.\,\ref{fig:diff_vac_SN_PC}), the LGBs produced on the far side of the shell ($\sim90^\circ\text{--}270^\circ$), which travel through the core, have the escape probabilities comparable to the ones produced on the near side. 
Thus, the far side of the shell contributes only slightly less than the near side, due to which the glowing shell maintains the shape even for larger couplings, unlike BMPs 3 and 5.
A similar pattern of energy transport can also be shown for the $L$-modes of all the BMPs, with appropriate couplings, as their production/reabsorption is also dominated by PC/inverse-PC (decay) processes.


On the other hand, for BMPs 3 (middle-row) and 5 (bottom-row), the shell glows brighter at first due to production through the SC process and suppressed decay rate outside the core due to light $\mdp$ (recall, $\Gamma^{\rm PC}\propto g'^2\mdp^2$). As the coupling is increased, the glowing shell changes shape. This is because the LGBs produced at angular locations $\sim 120^\circ$–$140^\circ$ and $\sim 220^\circ$–$240^\circ$ experience maximal attenuation as they traverse the longest path through the high-temperature, muon-rich shell where the inverse SC rate is large. 
In contrast, LGBs originating near $\sim 180^\circ$ propagate almost perpendicular to the shell and spend relatively less time in the region, resulting in comparatively weaker attenuation.
At even larger couplings, the LGBs produced on the far side get largely reabsorbed while travelling through the core. Consequently, the glowing shell gradually reduces to an arc on the side facing $A_{\rm unit}$. At this stage, upon integrating over the entire surface of the $R_{\rm far}$-sphere, the contribution only arrives from a thin shell, signifying the formation of \emph{LGB-sphere} and its surface emissions. Notably, the arc-like feature appears at smaller couplings for BMP \textbf{3} compared to BMP \textbf{5}. This is because, outside the \emph{LGB-sphere}, absorption is dominated by decay (inverse PC) processes, whose rate is larger for BMP \textbf{3} due to its higher LGB mass, as explained earlier.

\begin{figure}[htb!]
	\centering
	\includegraphics[width=0.8\linewidth]{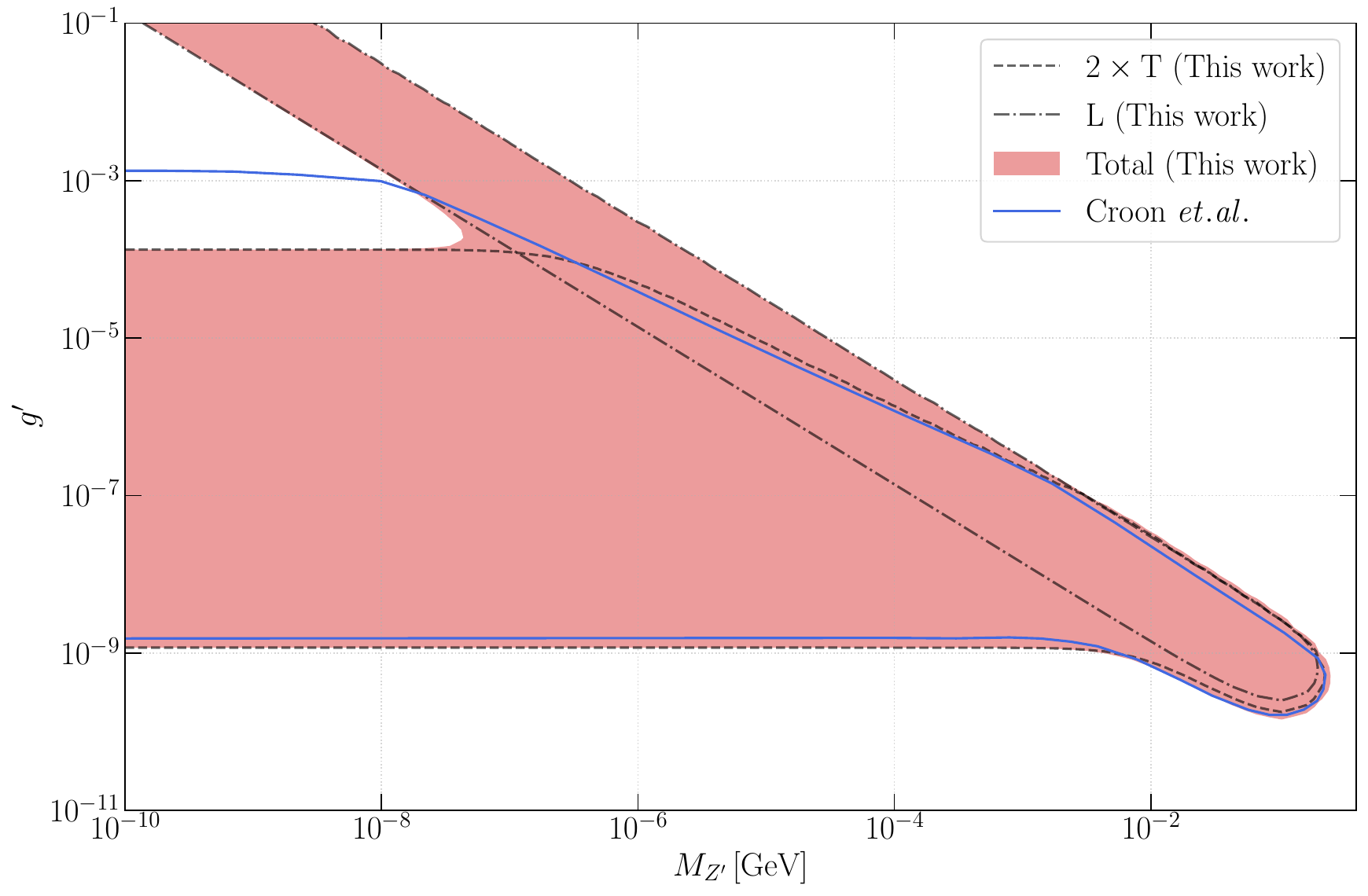}
	\caption{The excluded region in the $\mdp$-$g'$ plane for the $L_\mu-L_\tau$ model, obtained from SN cooling constraints due to LGBs, is shown in light-red for $R_{\rm far}=100$\,km. The dashed and dot-dashed contours correspond to Raffelt's limit for the combined two-transverse modes and the longitudinal mode, respectively. The blue curve show the limits evaluated in Croon \emph{et.al.}~\cite{Croon:2020lrf}. 
    The white patch enclosed by the light-red region remains unconstrained, as the combined transverses and longitudinal contributions do not exceed Raffelt's limit. 
    In this region, the transverse modes are strongly trapped, while the longitudinal modes are inefficiently produced.
	}
	\label{fig:excl}
\end{figure}

Summing up everything, we figure out the region in the parameter space that is disfavored from the observation of SN1987A. As discussed above, the region of parameter space in which the luminosity due to LGB violates Raffelt’s criterion is excluded \cite{Raffelt:1996wa}.
Fig.\,\ref{fig:excl} shows the excluded region in the $\mdp$-$g'$ plane colored in light-red. 
The region enclosed by the dashed and dot-dashed curves corresponds to the parameter space in which the luminosities of transverse ($2 L_{Z'}^T$) and longitudinal ($L_{Z'}^L$) modes, respectively, exceed Raffelt's criterion luminosity. The solid blue curve shows the limits as evaluated in Croon {\it et.al.}~\cite{Croon:2020lrf}\footnote{Restricting our analysis to the transverse mode only, we obtain results similar to those of Croon {\it et al.}~\cite{Croon:2020lrf} when adopting $R_{\rm far}=100$\,km and including only the SC and PC/decay channels. However, a small difference in the trapping regime of the $T$-mode appears at low masses, due to the additional attenuation arising from inverse-LB processes in our analysis. The bounds reported in Lai \textit{et al.}~\cite{Lai:2024mse}, which include LB processes and take $R_{\rm far}=R_\nu$, are likewise expected to be modified when the longitudinal and transverse polarisations are treated separately.
}. 
The region below the constraints is the region of insufficient production, while the region above is the large trapping region of the parameter space. The \emph{white~patch} enclosed by the light-red region in the top-left of the plot remains unconstrained in our studies, as in this region, the transverse modes are highly trapped, whereas the longitudinal modes are not efficiently produced. This is one of the key highlighted features of the independent treatment of the polarisation, which can be understood from the above Fig.\,\ref{fig:lmutauExcl} for the light mass and higher coupling regime (see, for example, BMP {\bf 6}).

\begin{figure}[htb!]
	\centering
	\includegraphics[width=0.8\linewidth]{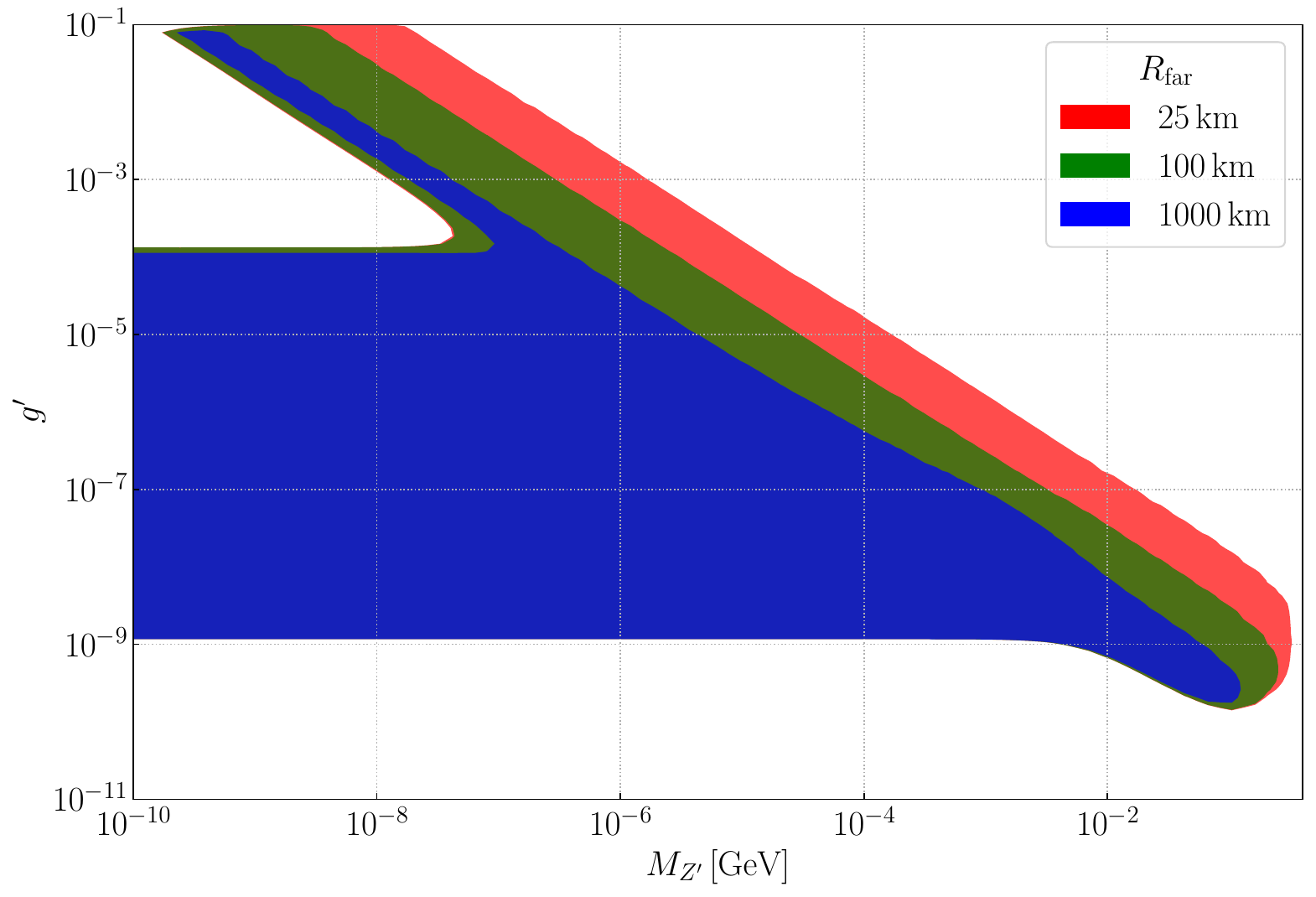}
	\caption{Excluded region for three different values of $R_{\rm far}$ as mentioned in Eq.\eqref{eq:rfars}, in different colors.}
	\label{fig:exclRfars}
\end{figure}

To assess the robustness of our results with respect to the choice of $R_{\rm far}$, we determine the excluded regions for three representative values,
\begin{align}
	R_{\rm far}\in\{R_\nu+5\,{\rm km}, R_{\rm gain},R_{\rm shock}\} .
	\label{eq:rfars}
\end{align}
Here, $R_\nu = 20\,\mathrm{km}$ denotes the neutrino-sphere radius, $R_{\rm gain} = 100\,\mathrm{km}$ is the gain radius beyond which neutrino production becomes inefficient, and $R_{\rm shock} = 1000\,\mathrm{km}$ corresponds to the shock radius, beyond which the material has not yet been shock heated~\cite{Chang:2016ntp}. The corresponding exclusion limits are shown in Fig.\,\ref{fig:exclRfars}.
Outside the neutrino sphere, the only process that attenuates the LGB flux produced in the core is the vacuum decay $Z' \to \nu\bar\nu$. For $R_{\rm far} = 25\,\mathrm{km}$, \emph{i.e.} when the energy carried by the LGB is considered lost once it crosses this radius, the escape distance is relatively short. Consequently, the reduction in luminosity due to decay is smaller than in the cases $R_{\rm far} = 100\,\mathrm{km}$ or $1000\,\mathrm{km}$, where the LGBs propagate over a longer distance and therefore have a greater probability to decay before escaping the supernova environment (see Fig.\,\ref{fig:snegeometry}). Therefore, the limits in the trapping regime relax for larger $R_{\rm far}$.

\section{Summary and Conclusions}\label{sec:SnC}
The observation of the supernova explosion SN1987A provides a unique astrophysical laboratory for testing new weakly coupled long-lived particles, which are expected to be produced in the hot, dense core of the star and escape, providing a new heat sink. This conflicts with the observed neutrino signal from SN1987A, which suggests that most of the energy is carried out of the PNS core through neutrinos, providing constraints on parameter space. In this work, we provide a comprehensive analysis of the effects of LGBs on SN1987A, treating the polarisation contributions to the luminosity ($L$-mode and $2~T$-modes) independently. This independent treatment leads to significant changes in the energy transportation dynamics, and hence in the exclusion in the parameter space compared to the previous studies.  

For illustration, we have adopted the simple $U(1)_{L_\mu-L_\tau}$ gauge extended scenario, in which supernova muons and neutrinos ($\nu_{\mu,\tau}$) couple to the LGB ($Z'$). In addition, protons in the supernova may couple to $Z'$ via loop-induced processes. In this scenario,   \emph{Neutrino pair coalescence}, \emph{semi-Compton}, and \emph{loop bremsstrahlung} processes are considered, relevant for SN cooling, and their contributions to the energy loss are evaluated independently for each polarisation mode, assuming no mixing during their propagation. Although mixing between different polarisations ($L$ and $T$) can arise via {\it dark Compton} scattering, it remains suppressed in the small coupling regime, \textit{i.e. }$g'< 0.1$~. For the \emph{neutrino pair coalescence} process, we include the effects of neutrino degeneracy in the production and absorption rates, instead of vacuum rates, and find a small effect on the overall results. The attenuation factor of the absorption rates is evaluated numerically rather than approximated analytically, thus reducing uncertainties. To provide further insight, we identify regions of large and small contributions to the total luminosity as functions of $Z'$ mass and coupling, indicating production regions with minimal and maximal attenuation relative to the exit point.

The independent treatment of $Z'$ polarisations ($L$ and $2~T$) leads to the following findings. For transverse modes, \emph{pair coalescence} processes dominate at higher $M_{Z'}$, whereas \emph{semi-Compton} processes dominate at lower masses. \emph{Loop bremsstrahlung} has a negligible effect on the production rate but slightly reduces the mean free path of $Z'$ in the core, leading to a consistently subdominant contribution. For the longitudinal mode, the \emph{pair coalescence} process dominates the production rate across the entire mass range, with the luminosity peak shifting to higher couplings as $M_{Z'}$ decreases. These distinct features of each polarisation mode contributing to the energy loss in the supernova, expressed as the total luminosity $L_{Z'}=2~ L_{Z'}^T+ L_{Z'}^L$, must lie below Raffelt’s bound to be consistent with SN1987A observations. For higher masses ($M_{Z'} \gtrsim 1$ MeV), both the modes ($L$ and $T$) contribute comparably to the total luminosity.
In the low-mass regime ($M_{Z'} \lesssim 1$ keV), transverse modes dominate at low couplings but are suppressed at higher couplings due to trapping, where the longitudinal mode takes over. This feature becomes prominent in the very low-mass regime, $M_{Z'} \lesssim \mathcal{O}(10)$ eV, for couplings $10^{-4} \lesssim g' \lesssim 0.1$, where the transverse contribution is strongly suppressed due to trapping, while the longitudinal production remains insufficient, and the resulting energy loss remains below Raffelt's bound. This represents the key difference arising from the independent polarisation treatment, leading to significant modifications of the SN cooling exclusion region in the model parameter space.  We also find that the exclusion limits, especially in the trapping region, are sensitive to the choice of $R_{\rm far}$. Notably, such an independent polarisation treatment is equally applicable to other light bosons with non-zero spin in SNe and could affect the corresponding exclusion limits to some extent.

\section*{Acknowledgment}\label{sec:Ack}
The computation in this work was partially supported by SAMKHYA, the high-performance computing facility provided by the Institute of Physics, Bhubaneswar, India (IOPB). 
The authors thank A.K. Saha for contributions at the early stages of this work. The authors also acknowledge J.H. Chang, W. DeRocco, R. Essig, and S.D. McDermott for valuable discussions. PG gratefully acknowledges the support of IOPB, where this project was first initiated during his tenure as a Visiting Scientist.

\appendix
\section*{APPENDIX}
\section{Phase space integral for LGB Production through PC process}
\label{app:phasespace}
The production through neutrino pair-coalescence is modified due to the finite temperature and chemical potential of neutrinos inside the core. Given the distribution of neutrinos $f_{\nu}$ (antineutrinos $f_{\bar\nu}$), the phase space integral can be solved as, 
\begin{align}
    \Gamma^{\rm\lambda, PC}_{\rm prod}&=\frac{1}{2\omega}
    \int \frac{d^3\vec{p}_\nu}{(2\pi)^3}\frac{1}{2E_\nu}f_{\nu}(E_\nu)
    \int \frac{d^3\vec{p}_{\bar{\nu}}}{(2\pi)^3}~\frac{1}{2E_{\bar\nu}}f_{\bar{\nu}}(E_{\bar{\nu}})
    (2\pi)^4\delta^{(4)}(p_{Z'}-p_\nu-p_{\bar{\nu}})
    |\mathcal M_{\nu_L {\bar{\nu}_R}\to Z'_\lambda}|^2
    \Bigg|_{\scriptsize{\begin{aligned}
    E_{\bar{\nu}}&=|\vec{p}_{\bar{\nu}}|\\E_\nu&=|\vec{p}_\nu|
    \end{aligned}}}\nonumber\\
    &=\frac{1}{2\omega}
    \int \frac{d^3\vec{p}_\nu}{(2\pi)^2}\frac{1}{4E_\nu E_{\bar\nu}}f_{\nu}(E_\nu)f_{\bar{\nu}}(E_{\bar{\nu}})
    \delta(\omega-E_\nu-E_{\bar{\nu}})|\mathcal M_{\nu_L {\bar{\nu}_R}\to Z'_\lambda}|^2
    \Bigg|_{\scriptsize{\begin{aligned}E_{\bar{\nu}}&=\sqrt{|\vec{p}_{Z'}|^2+|\vec{p}_\nu|^2-2\vec{p}_{Z'}\cdot\vec{p}_\nu}\\E_\nu&=|\vec{p}_\nu|\end{aligned}}}\nonumber\\
    &~\begin{aligned}=\frac{1}{2\omega}
    \int d\cos\theta_\nu&\frac{dE_\nu ~}{2\pi}\frac{E_\nu}{\omega-E_\nu}f_{\nu}(E_\nu)f_{\bar{\nu}}\left(\omega-E_\nu\right)
    \\&\times\delta\left(\omega-E_\nu-\sqrt{\omega^2-\mdp^2+E_\nu^2-2E_\nu\sqrt{\omega^2-\mdp^2}\cos\theta_\nu}\right)|\mathcal M_{\nu_L {\bar{\nu}_R}\to Z'_\lambda}|^2\end{aligned}\nonumber\\
    &~\begin{aligned}=\frac{1}{16\pi\omega}
    \int d\cos\theta_\nu&~dE_\nu \frac{E_\nu}{\omega-E_\nu}f_{\nu}(E_\nu)f_{\bar{\nu}}\left(\omega-E_\nu\right)
    \\&\times\left|\frac{\omega-E_\nu}{\omega - \sqrt{\omega ^2-\mdp^2}\cos \theta_\nu}\right|\delta\left(E_\nu-\frac{\mdp^2}{2 \left(\omega -\cos \theta_\nu \sqrt{\omega ^2-\mdp^2}\right)}\right)|\mathcal M_{\nu_L {\bar{\nu}_R}\to Z'_\lambda}|^2\end{aligned}\nonumber\\
    &=\frac{1}{16\pi\omega}
    \int d\cos\theta_\nu\frac{2E_\nu^2}{\mdp^2}f_{\nu}(E_\nu)f_{\bar{\nu}}(\omega-E_\nu)
    |\mathcal M_{\nu_L {\bar{\nu}_R}\to Z'_\lambda}|^2\Bigg|_{E_\nu=\frac{\mdp^2}{2 \left(\omega -\cos \theta_\nu \sqrt{\omega ^2-\mdp^2}\right)}}\nonumber\\
\end{align}
where, ${p}_{\nu,\bar\nu}$, $\vec{p}_{\nu,\bar\nu}$ and $E_{\nu,\bar\nu}$ is the neutrino/antineutrino four-momentum, three-momentum and energy respectively, $\theta_\nu$ is the angle of incoming neutrino with respect to the direction of outgoing LGB and $\omega$ is the energy of the LGB. Substituting $\omega=\mdp \cosh\alpha$ into the solution of $E_\nu$ gives, 
\begin{align}
    E_\nu=\frac{\mdp}{2(\cosh\alpha-\cos\theta_\nu\sinh\alpha)}
\end{align}
where $\alpha$ is the rapidity of LGB, defined as $\tanh\alpha\equiv v_{Z'}=\sqrt{1-M_{Z'}^2/\omega^2}$
\bibliographystyle{utphys}
\bibliography{ref}
\end{document}